**Vagus nerve stimulation strengthens beta- and gamma-rate auditory steady-state responses in rat cortex**

Shinichi Kumagai[a,b], Tomoyo Isoguchi Shiramatsu[b], Karin Oshima[b], Yohei Ishishita[a], Kenji Ibayashi[a], Yoshiyuki Onuki[a], Naoto Kunii[a], Kensuke Kawai[a], Hirokazu Takahashi[b*]

[a]Department of Neurosurgery, Jichi Medical University, Tochigi, Japan; [b]Department of Mechano-Informatics, Graduate School of Information Science and Technology, The University of Tokyo, Tokyo, Japan

*Correspondence to: Hirokazu Takahashi, Ph.D, Department of Mechano-Informatics, Graduate School of Information Science and Technology, The University of Tokyo, 7-3-1 Hongo, Bunkyo, Tokyo 113-8656, Japan. Tel: +81-3-5841-0461. Email: takahashi@i.u-tokyo.ac.jp

Abbreviations: VNS, vagus nerve stimulation; GENUS, Gamma ENtrainment Using Sensory stimulation; ACcore, core auditory cortex; ACbelt, belt auditory cortex; IAF, insular auditory field; AIC, anterior insular cortex; ECoG, electrocorticography; AEP, auditory-evoked potential; MCA, middle cerebral artery; ITPC, inter-trial phase clustering; ERSP, event-related spectral perturbation; WPLI, debiased weighted phase-lag index; ASSR, auditory steady-state response.

**Abstract**

***Background:*** Gamma entrainment using sensory stimulation (GENUS) is under study as a treatment for neurological disease, but cortical engagement is variable. Vagus nerve stimulation (VNS) modulates cortical state and sound-evoked beta–gamma activity, suggesting that it may strengthen responses to rhythmic sensory input.

***Objective:*** To determine whether VNS strengthens auditory steady-state responses across beta and gamma rates and whether the effect depends on cortical region and sensory drive.

***Methods:*** Electrocorticography (ECoG) was recorded from core and belt auditory cortex, insular auditory field (IAF), and anterior insular cortex (AIC) in seven isoflurane-anesthetized rats before and 90 min after initiation of 30-Hz cervical VNS. Responses to 500-ms click trains at 20 and 40 Hz were quantified using inter-trial phase clustering (ITPC), event-related spectral perturbation (ERSP), and debiased weighted phase-lag index (WPLI).

***Results:*** VNS increased click-evoked P1–N1 amplitude. During 40-Hz trains, 40-Hz ITPC and ERSP increased only in auditory core. During 20-Hz trains, 20-Hz ITPC increased in core, belt, and IAF, whereas ERSP increased only in core; 40-Hz harmonic responses increased only in core. AIC showed no change. Within auditory cortex, WPLI increased selectively at frequencies present in each click train and not in the alpha band.

***Conclusions:*** VNS strengthened beta- and gamma-rate stimulus-locked responses in auditory-driven cortex without globally increasing synchrony. VNS may therefore enhance auditory cortical target engagement during GENUS, although effects on disease pathology or clinical outcomes remain untested.

Keywords: Vagus nerve stimulation, Entrainment, Sensory stimulation, Gamma oscillation, Beta oscillation, Auditory cortex, Insular cortex.

## Introduction

Rhythmic sensory stimulation can drive frequency-specific cortical activity and is being explored therapeutically across neurology and psychiatry [1, 2]. The best-developed example is 40-Hz gamma entrainment using sensory stimulation (GENUS), which reduces Alzheimer-related pathology and improves cognition in mouse models [3-7]. Human studies indicate that visual, auditory, and audiovisual GENUS is feasible and can engage distributed cortical networks [8, 9], but clinical benefits remain preliminary and inconsistent [10-13].

Variable cortical target engagement may contribute to these mixed outcomes. Alzheimer's disease disrupts gamma activity and parvalbumin-interneuron function [14-16], and stimulus locking attenuates as activity propagates through cortical and hippocampal networks [17]. In an open-label extension, sustained EEG entrainment was associated with less cognitive decline [18]. This association is not proof that stronger entrainment improves outcomes, but it motivates interventions that can enhance stimulus-locked cortical responses.

One candidate is vagus nerve stimulation (VNS), a clinically established neuromodulation therapy for epilepsy and depression and, when paired with rehabilitation, for post-stroke motor recovery [19-21]. Vagal afferent pathways recruit noradrenergic, cholinergic, and serotonergic systems that regulate cortical state and excitation–inhibition balance [22-31]. Cholinergic recruitment of fast-spiking interneurons provides a plausible route for modulating

beta–gamma dynamics [32, 33]. Consistent with this mechanism, VNS alters beta–gamma synchrony in patients [34, 35] and amplifies sound-evoked beta–gamma power in rat auditory cortex through cholinergic mechanisms [36]. It remains unknown whether VNS strengthens stimulus-locked responses at the beta and gamma rates used in rhythmic sensory stimulation, and to what extent such modulation varies across cortical regions with different auditory response properties.

We addressed this question across four cortical regions: core auditory cortex (ACcore), surrounding belt (ACbelt), insular auditory field (IAF), and anterior insular cortex (AIC). IAF receives direct auditory thalamic input and provides an auditory-responsive pathway outside ACcore [37-42], whereas AIC lacks a clear stimulus-driven auditory response [37, 43-45]. This design allowed us to test whether any VNS effect tracks regional auditory drive rather than reflecting a generalized increase in cortical excitability.

Electrocorticography (ECoG) responses to 20- and 40-Hz click trains were measured before and after cervical VNS. We quantified the transient P1–N1 response, stimulus-locked phase consistency using inter-trial phase clustering (ITPC), spectral power using event-related spectral perturbation (ERSP), and non-zero-lag phase coupling between recording sites using the debiased weighted phase-lag index (WPLI). We predicted that VNS would strengthen beta- and gamma-rate responses primarily in auditory-driven regions and at frequencies represented in the stimulus.

## Material and Methods

### *Animals*

Procedures followed the ARRIVE (Animal Research: Reporting of In Vivo Experiments) guidelines and the Japanese Physiological Society's "Guiding Principles for the Care and Use of Animals in the Field of Physiological Science" and were approved by the animal ethics committee of the University of Tokyo's Graduate School of Information Science and Technology (JA22-7). Ten male Wistar rats (200–300 g, 8–12 postnatal weeks) were used. One rat was excluded because it emerged from anesthesia during recording, and two were excluded because ACcore could not be delineated, leaving seven rats. This sample size was based on previous studies using similar experimental preparations [46, 47]. All animals and electrophysiological recordings were newly acquired and independent of our previous study [36].

### *Surgical procedures*

Surgical procedures followed previous reports [36, 47]. Under isoflurane anesthesia (5% for induction and 1.5–2.5% for maintenance), a VNS system (Model 103, Cyberonics) was implanted at least five days before recording, with the generator subcutaneous in the back and a spiral electrode on the left vagus nerve. Electrophysiological experiments were also performed under isoflurane anesthesia. The right auditory and insular cortices were exposed by craniotomy and dural removal. Body temperature was maintained at 35–38°C, with continuous respiration and heart-rate monitoring. Artificial cerebrospinal fluid (ARTCEREB, Otsuka Pharmaceutical) and a neurosurgical pad (BEMSHEETS, Kawamoto) prevented cortical dehydration.

***VNS protocol***

VNS comprised 500-µA biphasic pulses at 30 Hz with a 130-µs pulse width, delivered in 900-pulse trains (30 s ON, 5 min OFF). Baseline recordings were obtained with the stimulator off. The ON/OFF cycle was then initiated and continued throughout subsequent recordings. Post-VNS recordings began 90 min after VNS onset (Fig. 1A), based on the reported time course of auditory cortical modulation [36].

***Auditory stimulation protocols***

Click trains are widely used to study auditory steady-state responses (ASSRs) [48] and elicit distinct responses [49]. Each click was a 0.2-ms rectangular pulse. For the 1-Hz click condition, sound pressure level was 100 dB SPL, measured using a microphone calibrated with a 1-kHz, 94-dB SPL reference tone. In the pre- and post-VNS conditions, 1-Hz clicks were presented for 15 and 60 min, respectively. Only the first 15 min (900 trials) of the post-VNS session were analyzed to match the pre-VNS condition. The 1-Hz click sessions were followed by 20-Hz and then 40-Hz sessions in fixed order, with 300 trials each. The two click-train sessions were separated by a 5-min silent interval. Each train lasted 500 ms, with fixed inter-click intervals and a 500-ms intertrain interval. A speaker (DLS-108X, Alpine Electronics) was positioned 15 cm from the left ear.

***Electrophysiological recordings***

Two 64-contact platinum surface arrays (NeuroNexus and Photo Precision; respective contact diameters, 100 and 80 µm; spacing, 500 µm) recorded epipially over auditory and insular cortex. The reference electrode was on the dura above contralateral parietal cortex. Auditory-evoked potentials (AEPs) were acquired at 1 kHz with a gain of 1000 and a 0.3–500-Hz digital bandpass filter (Cerebus, Blackrock Microsystems).

***Delineation of auditory and insular fields***

The ACcore and IAF were identified based on the spatial distribution of click-evoked AEPs obtained during the pre-VNS 1-Hz click session. Previous studies have shown that the core region, encompassing A1 and the anterior auditory field, exhibits larger AEPs than the surrounding areas [50-52], whereas the IAF is located approximately 2.7 mm anterior and 1.5 mm lateral to the anterior auditory field and comprises a responsive area with a diameter of about 2 mm [40]. Recording sites with absent signal or evident artifacts were excluded. For the auditory cortical array, the P1 amplitude at each recording site was z-scored, and the probability density of the z scores was estimated using a Gaussian kernel with the bandwidth given by the robust form of Silverman's rule. To distinguish the high-amplitude response population from the lower-amplitude population, the second derivative of the estimated density was computed. Its minimum was used as a reference point for identifying the first local maximum toward higher amplitudes, and the corresponding z score was defined as the threshold [53]. Sites exceeding this threshold were assigned to the ACcore (Supplementary Fig. S1). Rats in which the suprathreshold sites did not form a single contiguous cluster comprising at least eight electrodes (2 $mm^2$, approximately the mean area of A1) were excluded [54]. While this delineation might not include all of the core region, it captured the sites with the most distinct AEPs. Recording sites located ventral or posterior to the border of the ACcore were categorized as the ACbelt [51]. The IAF was defined as the contiguous electrodes whose P1 amplitude exceeded 50% of the maximum within the insular array [40]. In rodents, the insular cortex extends rostrocaudally along the rhinal vein [55], and the array was positioned parallel to this vessel. The AIC was defined as the recording sites in the two most rostral columns of the array, spanning three rows centred on the IAF peak. The number of electrodes assigned to each field varied across rats (ACcore, 9–25; ACbelt, 4–25; IAF, 2–7; AIC, 5–6; Supplementary Table S1).

***Evoked potential analysis***

Analyses used MATLAB R2022a (MathWorks). Consistent with prior recordings [40, 50], P1 was the positive peak within 20 ms of click onset and N1 the negative peak at 20–50 ms.

P1–N1 was their peak-to-peak difference. Amplitudes were normalized to each animal's maximum pre-VNS P1 across all analyzed sites.

***Time–frequency analysis***

To examine VNS-induced changes in auditory entrainment, the time–frequency representation of the ECoG signals was computed for each channel using complex Morlet wavelets, from which ITPC and ERSP were derived. Single-trial epochs were extracted from −900 to +900 ms relative to click onset. Trials containing samples exceeding 5 standard deviations of the corresponding channel signal were rejected as artifacts. The signals (sampled at 1 kHz) were convolved in the frequency domain with wavelets at 75 logarithmically spaced frequencies between 1 and 150 Hz, each with a fixed width of 7 cycles. ITPC was calculated at each time–frequency point as the length of the mean resultant vector across trials, yielding values from 0 (random phase) to 1 (perfect phase alignment). For ITPC, the number of trials was matched between the pre- and post-VNS conditions separately for each channel by random selection from the condition with more surviving trials. For ERSP, single-trial power (squared magnitude of the convolution output) was converted to decibels relative to the mean power of the −200 to −100 ms pre-stimulus baseline of each trial and then averaged across trials. For the 20-Hz, 40-Hz and alpha bands, ITPC and ERSP were averaged across the wavelet frequencies within 18–22, 38–42 and 8–13 Hz, respectively, and across 100–500 ms after train onset.

***Inter-electrode phase synchronization analysis***

Unlike ITPC, which indexes phase consistency relative to the stimulus, this analysis indexes the consistency of the phase difference between two electrodes. This was quantified from the same wavelet decomposition. Because volume conduction and the shared reference introduce a common, zero-lag component into neighboring contacts [56, 57], a local Laplacian derivation similar to the Hjorth method was applied [58]. Each contact was re-referenced to the mean of its artifact-free neighbors within two grid units (1.0 mm) on the same array. Contacts with zero variance in either recording were excluded. The same three bands defined above were analyzed (40, 20 Hz and alpha). Theta was excluded because the wavelet duration exceeds the analysis window. Trials were rejected using the same amplitude criterion as above, applied over the analysis window extended by three standard deviations of the Gaussian envelope of the wavelet at the lowest frequency of each band. Only trials artifact-free in every contact of the four regions of interest were retained, with trial number matched between conditions by random selection from the condition with more surviving trials.

For each contact pair, phase coupling was computed from the cross-spectrum of the two Laplacian-derived signals using the debiased estimator of the squared WPLI, which discounts zero-lag contributions [59]. WPLI was evaluated across trials at each time point of the 100–500 ms window and averaged over time points. Connectivity maps of WPLI were aligned across animals using the electrode closest to the P1-weighted centroid of the ACcore as a reference point, as described previously [36].

Surrogate distributions were generated by an independent random circular shift (10–90% of the analysis window) of each trial of each contact [60, 61]. Surrogate values from the pre- and post-VNS conditions, which had matched trial numbers, were pooled to form a single null distribution. On each of 1,000 iterations (500 per condition), the maximum WPLI across all contact pairs was retained, and the 95th percentile of this distribution defined a single significance threshold applied to all contact pairs. Thresholds were computed separately for each rat, click rate, and frequency.

***Statistical analysis***

P1–N1 amplitude, logit-transformed ITPC, and ERSP were analyzed using restricted-maximum-likelihood linear mixed models with Satterthwaite denominator degrees of freedom (SPSS Statistics v31, IBM), fitted separately by measure and, for ITPC and ERSP, by click-train condition and response band. Fixed effects were Region, Timing (pre/post), and their interaction. Random effects were rat and rat-nested electrode intercepts, with a Timing slope for region within rat. Benjamini–Hochberg correction was applied separately to pairwise regional comparisons and within-region Timing effects within each measure, click-train condition, and band.

For phase synchronization, the unit of analysis was the electrode pair (Supplementary

Table S1). WPLI was analyzed untransformed in three families defined by array: within the auditory array, across arrays, and within the insular array. Models (MATLAB R2022a, fitlme) replaced Region with Pair type, included centered inter-electrode distance for within-array families, and added intercepts for each electrode of a pair. Simple effects of Timing were estimated for each pair type regardless of the interaction and corrected within each family and band. The proportion of pairs exceeding a surrogate threshold was computed per rat within each band, logit-transformed, and compared between pre and post with paired *t* tests ($n = 7$), corrected across pair types within each family and band. To avoid infinite logits, proportions were adjusted to $(k + 0.5)/(n + 1)$ before logit transformation, where $k$ is the number of pair–frequency combinations exceeding the threshold and $n$ is their total number [62].

Frequency specificity was assessed by combining click rates in a single model per measure, classifying bands as driven (40 Hz during 40-Hz trains; 20 and 40 Hz during 20-Hz trains) or non-driven (all others, including alpha). Fixed effects were Driven status, Timing, their interaction, and Region (ITPC/ERSP) or Pair type and centered inter-electrode distance (WPLI; auditory array only). The 20-Hz band was also analyzed alone with click rate replacing Driven status. Detailed model specifications are provided in the Supplementary Methods.

## Results

### *Click-evoked responses in the auditory and insular cortices*

Approximately one week after implantation, ACcore and IAF showed distinct P1 peaks at 10–15 ms followed by N1, consistent with previous reports [40, 51-53]. IAF lay anteroventral to auditory cortex, dorsal to the rhinal vein and posterior to the MCA (Fig. 1B). ACbelt and AIC lacked a comparable onset response (Fig. 1C).

### *Effects of VNS on click-evoked P1–N1 amplitude*

Normalized P1–N1 amplitude showed main effects of Region ($F(3, 33.2) = 32.96$, $p < 0.001$) and Timing ($F(1, 22.4) = 13.16$, $p = 0.001$), without a Region × Timing interaction ($F(3, 22.4) = 1.76$, $p = 0.183$; Fig. 1D). Across regions, estimated marginal means increased from 0.47 to 0.77 (difference = 0.30). ACcore (1.19) exceeded IAF (0.89; $p = 0.034$), and both exceeded ACbelt (0.31) and AIC (0.08; all $p < 0.001$). ACbelt and AIC did not differ ($p = 0.084$). Thus, amplitude increased overall, without evidence for regional differences in the increase.

### *VNS enhances steady-state responses to periodic click trains*

ITPC and ERSP were analyzed at 40 Hz during 40-Hz trains and at 20 Hz and its 40-Hz harmonic during 20-Hz trains. All three conditions showed Region and Timing main effects for both measures (all $p \leq 0.003$); we therefore focus on interactions and within-region Timing effects.

During 40-Hz trains, ITPC and ERSP increases were concentrated in sustained ACcore responses (Fig. 2A, B; pre/post maps, Supplementary Fig. S2). The transient broadband onset response was similar before and after VNS; traces diverged thereafter through train offset (Fig. 2C, D). Other regions showed little change. To minimize the contribution of transient onset responses [63, 64], quantification was restricted to 100–500 ms to emphasize sustained responses.

Region × Timing interactions were significant for ITPC ($F(3, 26.5) = 4.71$, $p = 0.009$) and ERSP ($F(3, 30.7) = 5.40$, $p = 0.004$). ACcore ITPC increased from 0.134 to 0.284 (Δlogit = 0.94, 95% CI [0.56, 1.32], $p < 0.001$), and ERSP from −1.32 to 0.58 dB (Δ = 1.90, 95% CI [1.16, 2.64], $p < 0.001$). No other region changed significantly (ITPC, all $p \geq 0.199$; ERSP, all $p \geq 0.741$; Fig. 2E).

***Enhanced phase locking extends beyond ACcore during 20-Hz trains***

Twenty-hertz trains showed Region × Timing interactions for ITPC ($F(3, 25.8) = 8.14$, $p < 0.001$) and ERSP ($F(3, 26.9) = 5.23$, $p = 0.006$), with the same onset-versus-sustained pattern (Fig. 3A–D). ERSP increased significantly only in ACcore (−0.49 to 1.55 dB; Δ = 2.04, 95% CI [1.26, 2.83], $p < 0.001$; other regions, all $p \geq 0.352$). ITPC increased in ACcore (0.117 to 0.314; Δlogit = 1.24, 95% CI [0.84, 1.63], $p < 0.001$), IAF (0.091 to 0.164; Δlogit = 0.68, 95% CI [0.26, 1.10], $p = 0.005$), and ACbelt (0.081 to 0.119; Δlogit = 0.43, 95% CI [0.03, 0.82], $p = 0.049$), but not AIC ($p = 0.584$; Fig. 3E). Phase-locking enhancement thus extended more widely than power enhancement.

***The 40-Hz harmonic of the 20-Hz trains***

The 40-Hz harmonic also showed Region × Timing interactions (ITPC, $F(3, 26.8) = 4.51$, $p = 0.011$; ERSP, $F(3, 28.1) = 7.09$, $p = 0.001$). ACcore ITPC increased from 0.154 to 0.299 ($\Delta$logit = 0.85, 95% CI [0.50, 1.21], $p < 0.001$), and ERSP from −1.12 to 1.54 dB ($\Delta = 2.67$, 95% CI [1.78, 3.55], $p < 0.001$). Other regions showed no significant changes (all $p \geq 0.152$; Fig. 4). Significant 40-Hz enhancement was confined to ACcore at both train rates, consistent with response frequency influencing its spatial extent.

***Regional distribution of steady-state responses***

Across pre- and post-VNS conditions, ITPC was highest in ACcore, intermediate in IAF, and lowest in ACbelt and AIC. ACcore exceeded all other regions during 20-Hz trains and at the 40-Hz harmonic (all $p \leq 0.033$), and exceeded ACbelt and AIC but not IAF during 40-Hz trains (both $p < 0.001$; vs IAF, p = 0.058). IAF exceeded ACbelt and AIC during 40-Hz trains ($p = 0.010$ and 0.019) and at the 40-Hz harmonic ($p = 0.001$ and $p < 0.001$), and exceeded AIC but not ACbelt during 20-Hz trains ($p = 0.002$ and 0.130). ACbelt and AIC did not differ (all $p \geq 0.057$). ERSP, by contrast, was highest in ACcore in all conditions (all $p \leq 0.002$) and did not differ among the other three regions (all $p \geq 0.205$).

### *VNS strengthens phase coupling within the auditory cortex at the driven frequencies*

Auditory-array WPLI increases concentrated at driven frequencies (Fig. 5A; pre/post maps, Supplementary Fig. S3). During 40-Hz trains, 40-Hz WPLI showed a Timing effect ($F(1, 18.7) = 5.67$, $p = 0.028$), without Pair type or interaction effects ($p = 0.294$ and 0.395). The largest estimated increase, in ACcore–ACcore pairs ($\Delta = 0.069$, 95% CI [0.012, 0.126]), did not survive correction across pair types ($p = 0.063$; uncorrected $p = 0.021$; Fig. 5B). During 20-Hz trains, Timing effects occurred at 20 Hz ($F(1, 18.8) = 36.15$, $p < 0.001$) and 40 Hz ($F(1, 19.0) = 8.07$, $p = 0.010$), without interactions ($p = 0.288$ and 0.173). At 20 Hz, WPLI increased in ACcore–ACcore ($\Delta = 0.114$, 95% CI [0.064, 0.164], $p < 0.001$), ACcore–ACbelt (0.077 [0.028, 0.127], $p = 0.006$), and ACbelt–ACbelt pairs (0.059 [0.008, 0.111], $p = 0.026$). At the harmonic, only ACcore–ACcore pairs reached significance (0.077 [0.027, 0.127], $p = 0.014$).

During 40-Hz trains, 20-Hz WPLI did not change ($F(1, 17.6) = 0.071$, $p = 0.793$). Alpha-band WPLI also showed no significant increase during either train rate ($p = 0.163$ and 0.076); all estimated changes were negative. This pattern argues against a generalized increase in synchrony.

Threshold-based analysis showed a similar pattern (Fig. 5C). The proportion of significantly coupled ACcore–ACcore pairs increased during 40-Hz trains (0.16 to 0.33, $p = 0.031$) and during 20-Hz trains at both 20 Hz (0.12 to 0.26, $p = 0.046$) and 40 Hz (0.19 to 0.33, $p$

= 0.034). ACcore–ACbelt pairs increased at 20 Hz during 20-Hz trains (0.06 to 0.16, $p = 0.011$). Non-driven bands showed no significant changes (all $p \geq 0.289$).

Among cross-array pairs, no pair type changed significantly at the driven frequencies (all $p \geq 0.393$). The only significant cross-array change was a decrease in alpha-band coupling of ACbelt–IAF pairs during 40-Hz trains ($\Delta = -0.054$, 95% CI [−0.094, −0.015], $p = 0.037$). Within the insular array, no pair type changed significantly at any frequency (all $p \geq 0.165$). These analyses are shown in Supplementary Fig. S4.

***Frequency specificity across the three measures***

Pooled models showed greater pre-to-post changes in driven than non-driven bands (Driven × Timing: ITPC, $F(1, 41.4) = 12.92$, $p < 0.001$; ERSP, $F(1, 41.1) = 14.48$, $p < 0.001$; WPLI, $F(1, 40.4) = 6.44$, $p = 0.015$). Restricting analysis to 20 Hz, driven during 20-Hz but not 40-Hz trains, the Click rate × Timing interaction was significant for ITPC ($F(1, 12.2) = 7.35$, $p = 0.019$) and ERSP ($F(1, 12.6) = 9.96$, $p = 0.008$), but not WPLI ($F(1, 12.0) = 4.02$, $p = 0.068$). For 20-Hz WPLI, the estimated pre-to-post change was 0.079 (95% CI [0.025, 0.133]) during 20-Hz trains and 0.009 (95% CI [−0.045, 0.063]) during 40-Hz trains.

## Discussion

VNS strengthened transient click-evoked potentials, sustained phase consistency and power, and auditory-cortical phase coupling. Significant steady-state enhancement was regionally selective: 40-Hz responses increased in ACcore, whereas 20-Hz phase consistency also increased in ACbelt and IAF. AIC showed no significant change, and coupling increased at stimulus-represented frequencies, supporting preferential facilitation of externally driven beta–gamma responses.

The effect was not gamma-specific. VNS increased responses at 20 Hz, 40 Hz, and the 40-Hz harmonic of 20-Hz trains. Prior work shows that cholinergic manipulations modulate ASSRs, particularly at 20 and 40 Hz [65, 66], and that fast-spiking interneurons support oscillatory activity across the 20–80-Hz range [33, 67, 68]. Together with evidence that VNS recruits cholinergic pathways [22-24, 69] and that cholinergic blockade prevents VNS-induced enhancement of sound-evoked beta–gamma power [36], these findings make cholinergic modulation a plausible mechanism, although it was not directly tested here.

Regional selectivity further constrains interpretation. Enhancement occurred in ACcore and, at 20 Hz, in ACbelt and IAF, but not in AIC. This pattern is consistent with dependence on an auditory-responsive substrate, but AIC differs from the other regions in laminar organization, cell composition, and baseline dynamics; sensory drive is therefore not established as a causal requirement. A plausible contributor is the granular, thalamorecipient circuitry that generates much of the ASSR [70, 71]. IAF occupies granular posterior insula and receives auditory thalamic input [41, 42], whereas AIC is agranular. The IAF result also extends VNS modulation of auditory processing beyond conventional auditory cortex.

Auditory-cortical phase coupling increased at 40 Hz during 40-Hz trains and at 20 and 40 Hz during 20-Hz trains. Pooled models supported greater changes in driven than non-driven bands. However, the direct comparison between train rates at 20 Hz did not reach significance for WPLI ($p = 0.068$). The coupling results thus support preferential facilitation of stimulus-driven responses, with less conclusive evidence for train-rate dependence within a fixed band.

These findings motivate testing VNS as an adjunct to auditory GENUS. Forty-hertz phase locking, power, and coupling increased with 30-Hz VNS, showing that frequency matching was unnecessary under the tested conditions. VNS may strengthen cortical target engagement when disease-related dysfunction limits auditory responses. Twenty-hertz effects suggest relevance to beta-rate protocols and disorders with impaired 20-Hz ASSRs [72]. Stronger target engagement, however, cannot be equated with therapeutic efficacy: these healthy anesthetized rats were assessed without pathological or behavioral outcomes.

Several limitations qualify the conclusions. Isoflurane suppresses rapid ASSRs [73, 74], and awake-state effects may differ. A concurrent sham group was not included. In a comparable preparation, our previous study found stable P1 amplitudes across time in control animals without VNS [36], and the present regional and frequency selectivity argues against a global time or anesthesia effect; nevertheless, time-dependent changes in ITPC, ERSP, and WPLI remain possible. Only 20- and 40-Hz trains were tested, so the boundary between the enhancement reported here and suppression at 6–8 Hz [75] is unknown. The proposed cholinergic mechanism was not manipulated directly. Finally, ITPC and ERSP cannot distinguish entrainment of an endogenous oscillator from repeated evoked responses [76, 77]; accordingly, the data support strengthened stimulus-locked or steady-state responses more directly than entrainment of intrinsic gamma oscillations.

**Conclusion**

VNS strengthened beta- and gamma-rate ASSRs in auditory-driven rat cortex and selectively increased coupling at frequencies represented in the stimulus. These results nominate VNS as an adjunct for increasing auditory cortical target engagement during GENUS. Whether this physiological enhancement improves disease pathology, cognition, or clinical outcomes requires direct testing.

**Funding**

This study was supported by JSPS KAKENHI (23H03023, 23H03465, 23KJ1866, 25H02600, 25K19944, 25K22825, 26H02517), AMED (24wm0625401h0001), JST (JPMJPR22S8), NEDO

(JPNP14004), the Asahi Glass Foundation, and the Secom Science and Technology Foundation. The funding agencies had no role in study design; data collection, analysis, or interpretation; writing of the report; or the decision to submit the article for publication.

**Declaration of competing interests**

None.

**Declaration of generative AI and AI-assisted technologies in the manuscript preparation process**

During the preparation of this work, the authors used Claude and ChatGPT to improve language, readability, and manuscript organization. After using these tools, the authors reviewed and edited the content as needed and take full responsibility for the content of the published article.

**CRediT authorship contribution statement**

SK: Conceptualization, Investigation, formal analysis, visualization, writing - original draft. TIS: Methodology, investigation, funding acquisition and writing - review and editing. KO: Investigation and writing - review and editing. YI: Methodology and writing - review and editing. KI: Methodology and writing - review and editing. YO: Methodology and writing - review and editing. NK: Supervision and writing - review and editing. KK: Supervision, writing - review and editing and funding acquisition. HT: Conceptualization, project administration, supervision, funding acquisition and writing - review and editing.

**Data statement**

The data supporting the findings of this study will be made publicly available in the Zenodo repository upon publication of this article.

**References**

[1] Thut G, Schyns PG, Gross J. Entrainment of perceptually relevant brain oscillations by non-invasive rhythmic stimulation of the human brain. Front Psychol 2011;2:170. https://doi.org/10.3389/fpsyg.2011.00170

[2] Adaikkan C, Tsai LH. Gamma Entrainment: Impact on Neurocircuits, Glia, and Therapeutic Opportunities. Trends Neurosci 2020;43(1):24-41. https://doi.org/10.1016/j.tins.2019.11.001

[3] Iaccarino HF, Singer AC, Martorell AJ, Rudenko A, Gao F, Gillingham TZ, et al. Gamma frequency entrainment attenuates amyloid load and modifies microglia. Nature 2016;540(7632):230-5. https://doi.org/10.1038/nature20587

[4] Martorell AJ, Paulson AL, Suk HJ, Abdurrob F, Drummond GT, Guan W, et al. Multi-sensory Gamma Stimulation Ameliorates Alzheimer's-Associated Pathology and Improves Cognition. Cell 2019;177(2):256-71.e22. https://doi.org/10.1016/j.cell.2019.02.014

[5] Adaikkan C, Middleton SJ, Marco A, Pao PC, Mathys H, Kim DN, et al. Gamma

Entrainment Binds Higher-Order Brain Regions and Offers Neuroprotection. Neuron 2019;102(5):929-43 e8. https://doi.org/10.1016/j.neuron.2019.04.011

[6] Park S-S, Park H-S, Kim C-J, Kang H-S, Kim D-H, Baek S-S, et al. Physical exercise during exposure to 40-Hz light flicker improves cognitive functions in the 3xTg mouse model of Alzheimer's disease. Alzheimer's Research & Therapy 2020;12(1):62. https://doi.org/10.1186/s13195-020-00631-4

[7] Murdock MH, Yang CY, Sun N, Pao PC, Blanco-Duque C, Kahn MC, et al. Multisensory gamma stimulation promotes glymphatic clearance of amyloid. Nature 2024;627(8002):149-56. https://doi.org/10.1038/s41586-024-07132-6

[8] He Q, Colon-Motas KM, Pybus AF, Piendel L, Seppa JK, Walker ML, et al. A feasibility trial of gamma sensory flicker for patients with prodromal Alzheimer's disease. Alzheimer's & Dementia: Translational Research & Clinical Interventions 2021;7(1). https://doi.org/10.1002/trc2.12178

[9] Chan DN, Suk HJ, Jackson BL, Milman NP, Stark D, Klerman EB, et al. Gamma frequency sensory stimulation in mild probable Alzheimer's dementia patients: Results of feasibility and pilot studies. Plos One 2022;17(12):e0278412. https://doi.org/10.1371/journal.pone.0278412

[10] Hajós M, Boasso A, Hempel E, Shpokayte M, Konisky A, Seshagiri CV, et al. Safety,

tolerability, and efficacy estimate of evoked gamma oscillation in mild to moderate Alzheimer's disease. Frontiers in Neurology 2024;15. https://doi.org/10.3389/fneur.2024.1343588

[11] Da X, Hempel E, Ou Y, Rowe OE, Malchano Z, Hajós M, et al. Noninvasive Gamma Sensory Stimulation May Reduce White Matter and Myelin Loss in Alzheimer's Disease. Journal of Alzheimer's Disease 2024;97(1):359-72. https://doi.org/10.3233/jad-230506

[12] Ismail R, Hansen AK, Parbo P, Brændgaard H, Gottrup H, Brooks DJ, et al. The Effect of 40-Hz Light Therapy on Amyloid Load in Patients with Prodromal and Clinical Alzheimer's Disease. International Journal of Alzheimer's Disease 2018;2018(1):1-5. https://doi.org/10.1155/2018/6852303

[13] Ang S, Zhang XY, Hong JL, Xue KY, Li JH, Du XZ, et al. The safety and efficacy of gamma frequency auditory and visual stimulation in the treatment of alzheimer's disease: a systematic review and meta-analysis. Translational Psychiatry 2025;15(1). https://doi.org/10.1038/s41398-025-03788-4

[14] Koenig T, Prichep L, Dierks T, Hubl D, Wahlund LO, John ER, et al. Decreased EEG synchronization in Alzheimer's disease and mild cognitive impairment. Neurobiology of Aging 2005;26(2):165-71. https://doi.org/10.1016/j.neurobiolaging.2004.03.008

[15] Verret L, Mann EO, Hang GB, Barth AM, Cobos I, Ho K, et al. Inhibitory interneuron deficit links altered network activity and cognitive dysfunction in Alzheimer model. Cell

2012;149(3):708-21. https://doi.org/10.1016/j.cell.2012.02.046

[16] Palop JJ, Mucke L. Network abnormalities and interneuron dysfunction in Alzheimer disease. Nat Rev Neurosci 2016;17(12):777-92. https://doi.org/10.1038/nrn.2016.141

[17] Schneider M, Tzanou A, Uran C, Vinck M. Cell-type-specific propagation of visual flicker. Cell Rep 2023;42(5):112492. https://doi.org/10.1016/j.celrep.2023.112492

[18] Chan D, de Weck G, Jackson BL, Suk HJ, Milman NP, Kitchener E, et al. Gamma sensory stimulation in mild Alzheimer's dementia: An open-label extension study. Alzheimers Dement 2025;21(10):e70792. https://doi.org/10.1002/alz.70792

[19] George M, Sonnen A, Upton A, Salinsky M, Ristanovic R, Bergen D, et al. A randomized controlled trial of chronic vagus nerve stimulation for treatment of medically intractable seizures. Neurology 1995;45(2):224-30. https://doi.org/10.1212/wnl.45.2.224

[20] Rush AJ, George MS, Sackeim HA, Marangell LB, Husain MM, Giller C, et al. Vagus nerve stimulation (VNS) for treatment-resistant depressions: a multicenter study. Biological Psychiatry 2000;47(4):276-86. https://doi.org/10.1016/s0006-3223(99)00304-2

[21] Dawson J, Liu CY, Francisco GE, Cramer SC, Wolf SL, Dixit A, et al. Vagus nerve stimulation paired with rehabilitation for upper limb motor function after ischaemic stroke (VNS-REHAB): a randomised, blinded, pivotal, device trial. Lancet 2021;397(10284):1545-53. https://doi.org/10.1016/S0140-6736(21)00475-X

[22] Hulsey DR, Hays SA, Khodaparast N, Ruiz A, Das P, Rennaker RL, 2nd, et al. Reorganization of Motor Cortex by Vagus Nerve Stimulation Requires Cholinergic Innervation. Brain Stimul 2016;9(2):174-81. https://doi.org/10.1016/j.brs.2015.12.007

[23] Collins L, Boddington L, Steffan PJ, McCormick D. Vagus nerve stimulation induces widespread cortical and behavioral activation. Curr Biol 2021;31(10):2088-98 e3. https://doi.org/10.1016/j.cub.2021.02.049

[24] Bowles S, Hickman J, Peng X, Williamson WR, Huang R, Washington K, et al. Vagus nerve stimulation drives selective circuit modulation through cholinergic reinforcement. Neuron 2022;110(17):2867-85 e7. https://doi.org/10.1016/j.neuron.2022.06.017

[25] Groves DA, Bowman EM, Brown VJ. Recordings from the rat locus coeruleus during acute vagal nerve stimulation in the anaesthetised rat. Neurosci Lett 2005;379(3):174-9. https://doi.org/10.1016/j.neulet.2004.12.055

[26] Hulsey DR, Riley JR, Loerwald KW, Rennaker RL, 2nd, Kilgard MP, Hays SA. Parametric characterization of neural activity in the locus coeruleus in response to vagus nerve stimulation. Exp Neurol 2017;289:21-30. https://doi.org/10.1016/j.expneurol.2016.12.005

[27] Dorr AE, Debonnel G. Effect of vagus nerve stimulation on serotonergic and noradrenergic transmission. J Pharmacol Exp Ther 2006;318(2):890-8. https://doi.org/10.1124/jpet.106.104166

[28] Hulsey DR, Shedd CM, Sarker SF, Kilgard MP, Hays SA. Norepinephrine and serotonin are required for vagus nerve stimulation directed cortical plasticity. Exp Neurol 2019;320:112975. https://doi.org/10.1016/j.expneurol.2019.112975

[29] Ruffoli R, Giorgi FS, Pizzanelli C, Murri L, Paparelli A, Fornai F. The chemical neuroanatomy of vagus nerve stimulation. J Chem Neuroanat 2011;42(4):288-96. https://doi.org/10.1016/j.jchemneu.2010.12.002

[30] Wang Y, Zhan G, Cai Z, Jiao B, Zhao Y, Li S, et al. Vagus nerve stimulation in brain diseases: Therapeutic applications and biological mechanisms. Neurosci Biobehav Rev 2021;127:37-53. https://doi.org/10.1016/j.neubiorev.2021.04.018

[31] Lee SH, Dan Y. Neuromodulation of brain states. Neuron 2012;76(1):209-22. https://doi.org/10.1016/j.neuron.2012.09.012

[32] Fisahn A, Pike FG, Buhl EH, Paulsen O. Cholinergic induction of network oscillations at 40 Hz in the hippocampus. Nature 1998;394(6689):186-9. https://doi.org/10.1038/28179

[33] Cardin JA, Carlen M, Meletis K, Knoblich U, Zhang F, Deisseroth K, et al. Driving fast-spiking cells induces gamma rhythm and controls sensory responses. Nature 2009;459(7247):663-7. https://doi.org/10.1038/nature08002

[34] Marrosu F, Santoni F, Puligheddu M, Barberini L, Maleci A, Ennas F, et al. Increase in 20–50Hz (gamma frequencies) power spectrum and synchronization after chronic vagal nerve

stimulation. Clinical Neurophysiology 2005;116(9):2026-36. https://doi.org/10.1016/j.clinph.2005.06.015

[35] Hyvarinen P, Yrttiaho S, Lehtimaki J, Ilmoniemi RJ, Makitie A, Ylikoski J, et al. Transcutaneous vagus nerve stimulation modulates tinnitus-related beta- and gamma-band activity. Ear Hear 2015;36(3):e76-85. https://doi.org/10.1097/AUD.0000000000000123

[36] Kumagai S, Shiramatsu TI, Matsumura A, Ishishita Y, Ibayashi K, Onuki Y, et al. Frequency-specific modulation of oscillatory activity in the rat auditory cortex by vagus nerve stimulation. Brain Stimul 2023;16(5):1476-85. https://doi.org/10.1016/j.brs.2023.09.019

[37] Zhang Y, Zhou W, Wang S, Zhou Q, Wang H, Zhang B, et al. The Roles of Subdivisions of Human Insula in Emotion Perception and Auditory Processing. Cereb Cortex 2019;29(2):517-28. https://doi.org/10.1093/cercor/bhx334

[38] Benarroch EE. Insular cortex: Functional complexity and clinical correlations. Neurology 2019;93(21):932-8. https://doi.org/10.1212/WNL.0000000000008525

[39] Evrard HC. The Organization of the Primate Insular Cortex. Front Neuroanat 2019;13:43. https://doi.org/10.3389/fnana.2019.00043

[40] Rodgers KM, Benison AM, Klein A, Barth DS. Auditory, somatosensory, and multisensory insular cortex in the rat. Cereb Cortex 2008;18(12):2941-51. https://doi.org/10.1093/cercor/bhn054

[41] Takemoto M, Hasegawa K, Nishimura M, Song WJ. The insular auditory field receives input from the lemniscal subdivision of the auditory thalamus in mice. J Comp Neurol 2014;522(6):1373-89. https://doi.org/10.1002/cne.23491

[42] Jankowski MM, Karayanni M, Harpaz M, Polterovich A, Nelken I. A Rapid Anterior Auditory Processing Stream through the Insulo-parietal Auditory Field in the Rat. The Journal of Neuroscience 2025;45(36):e2382242025. https://doi.org/10.1523/jneurosci.2382-24.2025

[43] Craig AD. How do you feel--now? The anterior insula and human awareness. Nat Rev Neurosci 2009;10(1):59-70. https://doi.org/10.1038/nrn2555

[44] Menon V, Uddin LQ. Saliency, switching, attention and control: a network model of insula function. Brain Struct Funct 2010;214(5-6):655-67. https://doi.org/10.1007/s00429-010-0262-0

[45] Craig AD. Significance of the insula for the evolution of human awareness of feelings from the body. Ann N Y Acad Sci 2011;1225(1):72-82. https://doi.org/10.1111/j.1749-6632.2011.05990.x

[46] Usami K, Kano R, Kawai K, Noda T, Shiramatsu TI, Saito N, et al. State-Dependent Modulation of Cortical Synchrony by Vagus Nerve Stimulation in Adult Rats. Electronics and Communications in Japan 2015;98(6):44-50. https://doi.org/10.1002/ecj.11735

[47] Takahashi H, Shiramatsu TI, Hitsuyu R, Ibayashi K, Kawai K. Vagus nerve stimulation

(VNS)-induced layer-specific modulation of evoked responses in the sensory cortex of rats. Sci Rep 2020;10(1):8932. https://doi.org/10.1038/s41598-020-65745-z

[48] O'Donnell BF, Vohs JL, Krishnan GP, Rass O, Hetrick WP, Morzorati SL. The auditory steady-state response (ASSR): a translational biomarker for schizophrenia. Application of Brain Oscillations in Neuropsychiatric Diseases - Selected Papers from "Brain Oscillations in Cognitive Impairment and Neurotransmitters" Conference, Istanbul, Turkey, 29 April–1 May 2011: Elsevier; 2013, p. 101-12.

[49] Tada M, Kirihara K, Koshiyama D, Fujioka M, Usui K, Uka T, et al. Gamma-Band Auditory Steady-State Response as a Neurophysiological Marker for Excitation and Inhibition Balance: A Review for Understanding Schizophrenia and Other Neuropsychiatric Disorders. Clin EEG Neurosci 2020;51(4):234-43. https://doi.org/10.1177/1550059419868872

[50] Takahashi H, Nakao M, Kaga K. Interfield differences in intensity and frequency representation of evoked potentials in rat auditory cortex. Hear Res 2005;210(1-2):9-23. https://doi.org/10.1016/j.heares.2005.05.014

[51] Shiramatsu TI, Kanzaki R, Takahashi H. Cortical mapping of mismatch negativity with deviance detection property in rat. PLoS One 2013;8(12):e82663. https://doi.org/10.1371/journal.pone.0082663

[52] Shiramatsu TI, Takahashi K, Noda T, Kanzaki R, Nakahara H, Takahashi H.

Microelectrode mapping of tonotopic, laminar, and field-specific organization of thalamo-cortical pathway in rat. Neuroscience 2016;332:38-52. https://doi.org/10.1016/j.neuroscience.2016.06.024

[53] Shiramatsu TI, Takahashi H. Mismatch Negativity in Rat Auditory Cortex Represents the Empirical Salience of Sounds. Front Neurosci 2018;12:924. https://doi.org/10.3389/fnins.2018.00924

[54] Kilgard MP, Pandya PK, Vazquez J, Gehi A, Schreiner CE, Merzenich MM. Sensory input directs spatial and temporal plasticity in primary auditory cortex. J Neurophysiol 2001;86(1):326-38. https://doi.org/10.1152/jn.2001.86.1.326

[55] Gogolla N. The insular cortex. Curr Biol 2017;27(12):R580-R6. https://doi.org/10.1016/j.cub.2017.05.010

[56] Nunez PL, Srinivasan R, Westdorp AF, Wijesinghe RS, Tucker DM, Silberstein RB, et al. EEG coherency. Electroencephalography and Clinical Neurophysiology 1997;103(5):499-515. https://doi.org/10.1016/s0013-4694(97)00066-7

[57] Shirhatti V, Borthakur A, Ray S. Effect of Reference Scheme on Power and Phase of the Local Field Potential. Neural Comput 2016;28(5):882-913. https://doi.org/10.1162/NECO_a_00827

[58] Hjorth B. An on-line transformation of EEG scalp potentials into orthogonal source

derivations. Electroencephalography and Clinical Neurophysiology 1975;39(5):526-30. https://doi.org/10.1016/0013-4694(75)90056-5

[59] Vinck M, Oostenveld R, van Wingerden M, Battaglia F, Pennartz CM. An improved index of phase-synchronization for electrophysiological data in the presence of volume-conduction, noise and sample-size bias. Neuroimage 2011;55(4):1548-65. https://doi.org/10.1016/j.neuroimage.2011.01.055

[60] Chen D, Kunz L, Lv PC, Zhang H, Zhou WJ, Liang SL, et al. Theta oscillations coordinate grid-like representations between ventromedial prefrontal and entorhinal cortex. Science Advances 2021;7(44):eabj0200. https://doi.org/10.1126/sciadv.abj0200

[61] Wu Y, Teng X, Du Y. Eye blinks synchronize with musical beats during music listening. PLoS Biol 2025;23(11):e3003456. https://doi.org/10.1371/journal.pbio.3003456

[62] Gart JJ, Zweifel JR. On the bias of various estimators of the logit and its variance with application to quantal bioassay. Biometrika 1967;54(1-2):181-7. https://doi.org/10.1093/biomet/54.1-2.181

[63] Wang X, Lu T, Snider RK, Liang L. Sustained firing in auditory cortex evoked by preferred stimuli. Nature 2005;435(7040):341-6. https://doi.org/10.1038/nature03565

[64] Marmelshtein A, Eckerling A, Hadad B, Ben-Eliyahu S, Nir Y. Sleep-like changes in neural processing emerge during sleep deprivation in early auditory cortex. Curr Biol

2023;33(14):2925-40 e6. https://doi.org/10.1016/j.cub.2023.06.022

[65] Zhang J, Ma L, Li W, Yang P, Qin L. Cholinergic modulation of auditory steady-state response in the auditory cortex of the freely moving rat. Neuroscience 2016;324:29-39. https://doi.org/10.1016/j.neuroscience.2016.03.006

[66] Sivarao DV, Frenkel M, Chen P, Healy FL, Lodge NJ, Zaczek R. MK-801 disrupts and nicotine augments 40 Hz auditory steady state responses in the auditory cortex of the urethane-anesthetized rat. Neuropharmacology 2013;73:1-9. https://doi.org/10.1016/j.neuropharm.2013.05.006

[67] Gulyás AI, Szabó GG, Ulbert I, Holderith N, Monyer H, Erdélyi F, et al. Parvalbumin-Containing Fast-Spiking Basket Cells Generate the Field Potential Oscillations Induced by Cholinergic Receptor Activation in the Hippocampus. The Journal of Neuroscience 2010;30(45):15134-45. https://doi.org/10.1523/jneurosci.4104-10.2010

[68] Pafundo DE, Miyamae T, Lewis DA, Gonzalez-Burgos G. Cholinergic modulation of neuronal excitability and recurrent excitation-inhibition in prefrontal cortex circuits: implications for gamma oscillations. The Journal of Physiology 2013;591(19):4725-48. https://doi.org/10.1113/jphysiol.2013.253823

[69] Kumagai S, Shiramatsu TI, Kawai K, Takahashi H. Vagus nerve stimulation as a predictive coding modulator that enhances feedforward over feedback transmission. Frontiers in

Neural Circuits 2025;19. https://doi.org/10.3389/fncir.2025.1568655

[70] Li Z, Li J, Wang S, Wang X, Chen J, Qin L. Laminar Profile of Auditory Steady-State Response in the Auditory Cortex of Awake Mice. Front Syst Neurosci 2021;15:636395. https://doi.org/10.3389/fnsys.2021.636395

[71] Li J, Li Z, Wang X, Liu Y, Wang S, Wang X, et al. The Thalamocortical Mechanism Underlying the Generation and Regulation of the Auditory Steady-State Responses in Awake Mice. J Neurosci 2024;44(1):e1166232023. https://doi.org/10.1523/JNEUROSCI.1166-23.2023

[72] Kinjo M, Tamura S, Sugiyama S, van Weert E, Oribe N, Tsuchimoto R, et al. A systematic review and meta-analysis of the auditory steady-state response in schizophrenia, bipolar disorder, and autism spectrum disorder. Mol Psychiatry 2026;31(6):3169-75. https://doi.org/10.1038/s41380-026-03452-3

[73] Plourde G, Villemure C, Fiset P, Bonhomme V, Backman SB. Effect of isoflurane on the auditory steady-state response and on consciousness in human volunteers. Anesthesiology 1998;89(4):844-51. https://doi.org/10.1097/00000542-199810000-00008

[74] Santarelli R, Carraro L, Conti G, Capello M, Plourde G, Arslan E. Effects of isoflurane on auditory middle latency (MLRs) and steady-state (SSRs) responses recorded from the temporal cortex of the rat. Brain Res 2003;973(2):240-51. https://doi.org/10.1016/s0006-8993(03)02520-4

[75] Nichols JA, Nichols AR, Smirnakis SM, Engineer ND, Kilgard MP, Atzori M. Vagus

nerve stimulation modulates cortical synchrony and excitability through the activation of muscarinic receptors. Neuroscience 2011;189:207-14. https://doi.org/10.1016/j.neuroscience.2011.05.024

[76] Duecker K, Gutteling TP, Herrmann CS, Jensen O. No Evidence for Entrainment: Endogenous Gamma Oscillations and Rhythmic Flicker Responses Coexist in Visual Cortex. Journal of Neuroscience 2021;41(31):6684-98. https://doi.org/10.1523/Jneurosci.3134-20.2021

[77] Soula M, Martin-Avila A, Zhang Y, Dhingra A, Nitzan N, Sadowski MJ, et al. Forty-hertz light stimulation does not entrain native gamma oscillations in Alzheimer's disease model mice. Nat Neurosci 2023;26(4):570-8. https://doi.org/10.1038/s41593-023-01270-2

## Figures

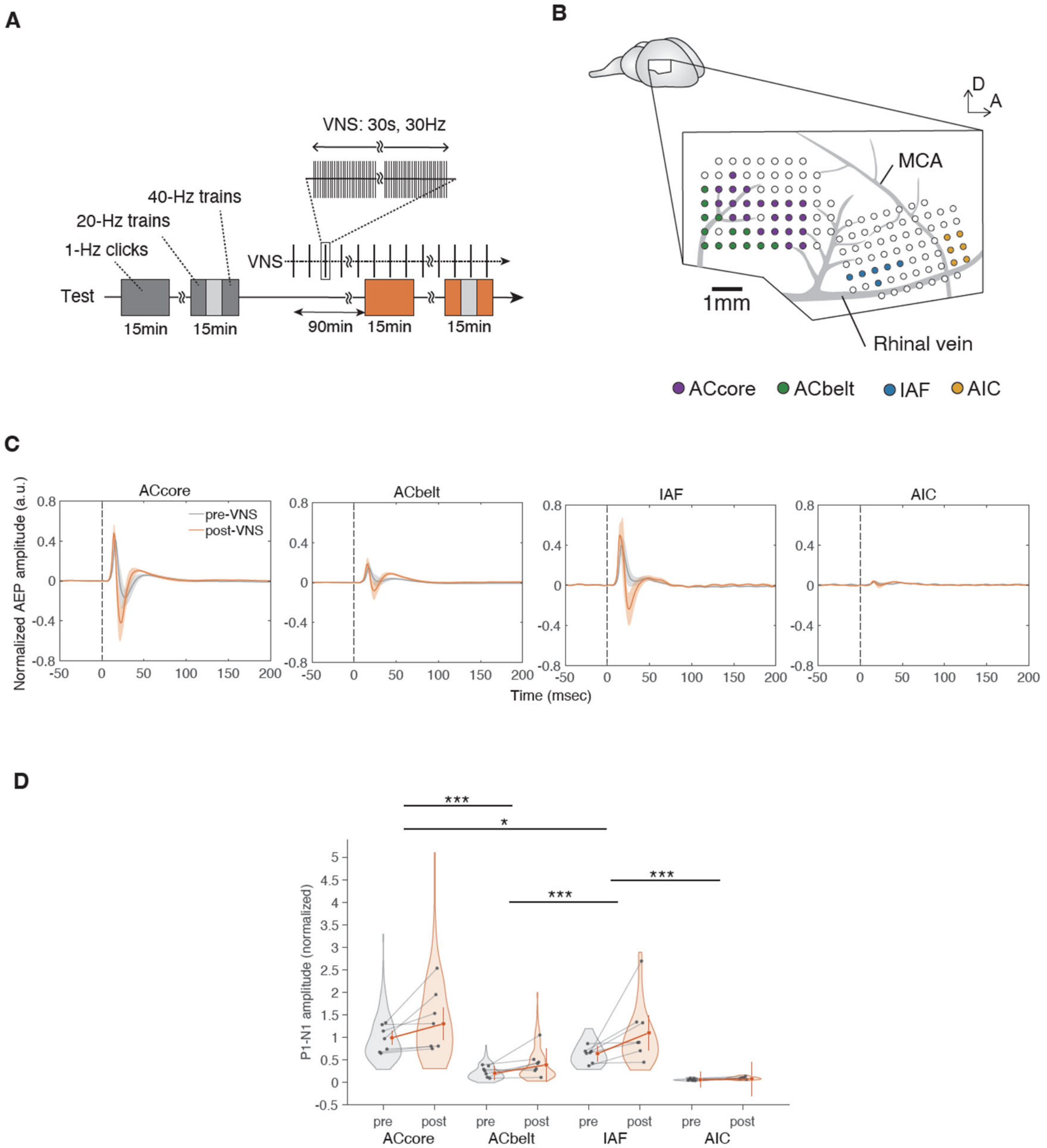

**Fig. 1** Electrode placement and effects of VNS on click-evoked responses. (A) Experimental timeline of the analyzed test sessions. Before (gray) and after (orange) VNS, single clicks at 1 Hz for 15 min were followed by 500-ms click trains at 20 Hz and then at 40 Hz for 5 min each, separated by a 5-min silent interval. (B) Representative electrode placement over the right

auditory and insular cortices. Filled circles indicate electrodes assigned to each field (purple, ACcore; green, ACbelt; blue, IAF; yellow, AIC); open circles were excluded. Scale bar, 1 mm. (C) Grand-average normalized AEP waveforms in each region before (gray) and after (orange) VNS. Shading indicates SEM. Dashed line, click onset. (D) Normalized P1–N1 amplitude before and after VNS in each region. Violin plots show the distribution across electrodes. Black dots represent individual animals and connecting lines indicate paired data. Orange circles and error bars indicate estimated marginal means with 95% confidence intervals from a linear mixed model. VNS increased P1–N1 amplitude across regions (main effect of Timing, $p = 0.001$), with no Region × Timing interaction. $n = 7$ rats. ***$p < 0.001$, *$p < 0.05$; n.s., not significant. VNS, vagus nerve stimulation; AEP, auditory evoked potential; ACcore, core auditory cortex; ACbelt, belt auditory cortex; IAF, insular auditory field; AIC, anterior insular cortex; MCA, middle cerebral artery.

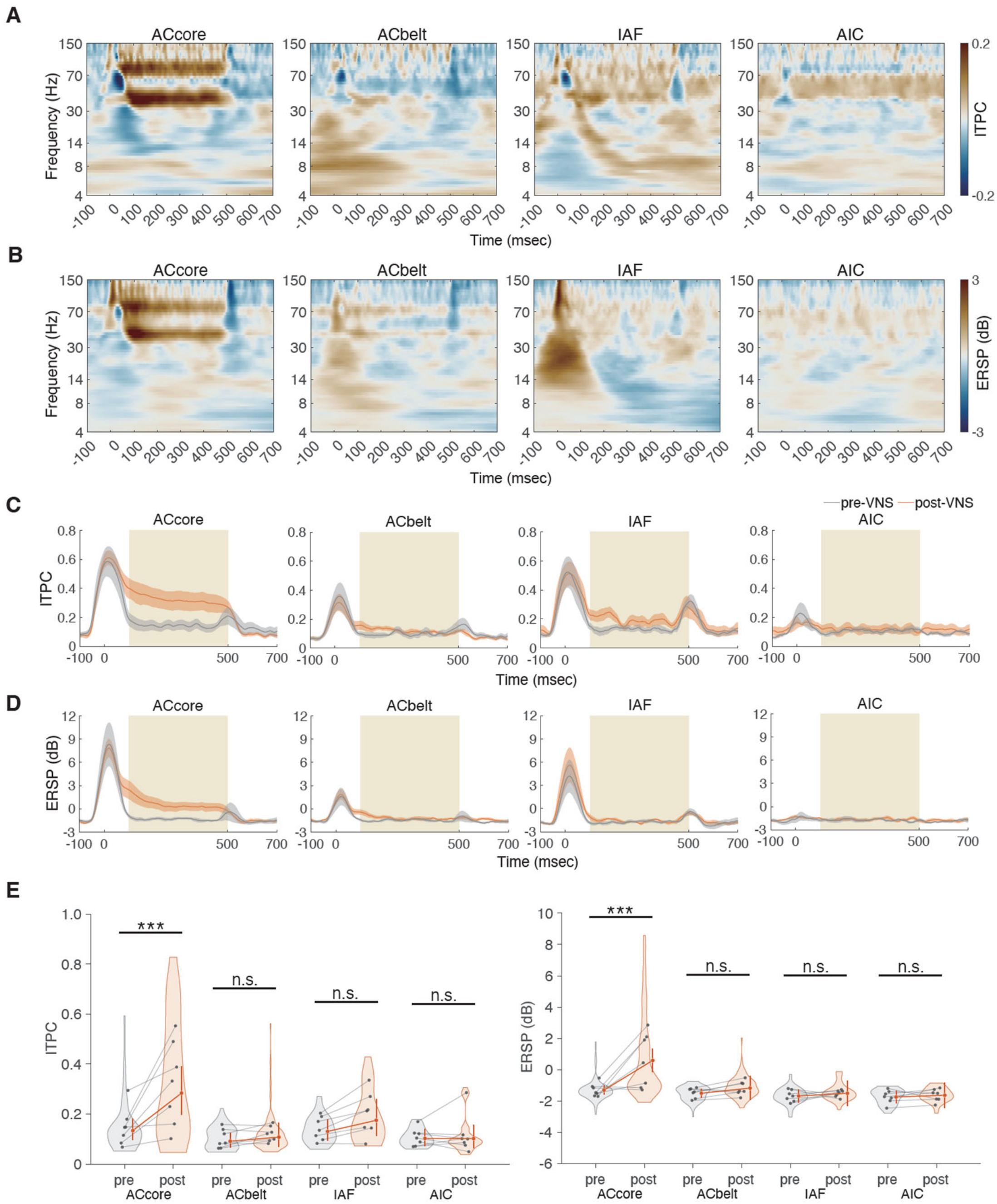

**Fig. 2** Effects of VNS on cortical entrainment to 40-Hz click trains. (A) Time–frequency maps of the VNS-induced change in ITPC (post − pre) for each region (ACcore, ACbelt, IAF, AIC). Time 0 indicates train onset and the train ended at 500 ms. Color scale indicates the change in ITPC on the original scale. (B) Time–frequency maps of the VNS-induced change in ERSP, displayed as in (A). Color scale indicates the change in ERSP (dB). (C) Time course of 40-Hz ITPC in each region (ACcore, ACbelt, IAF, AIC) before (gray) and after (orange) VNS. Shading around each trace indicates the SEM across rats. The beige vertical band denotes the 100–500 ms quantification window. (D) Time course of 40-Hz ERSP, displayed as in (C). (E) 40-Hz ITPC (left) and 40-Hz ERSP (right) before and after VNS in each region. Violin plots show the distribution across electrodes; black dots represent individual animals and connecting lines indicate paired data. Orange circles and error bars indicate estimated marginal means with 95% confidence intervals from a linear mixed model. Statistics were computed on logit-transformed ITPC and are displayed here on the original scale. $n = 7$ rats. ***$p < 0.001$; n.s., not significant. ITPC, inter-trial phase clustering; ERSP, event-related spectral perturbation.

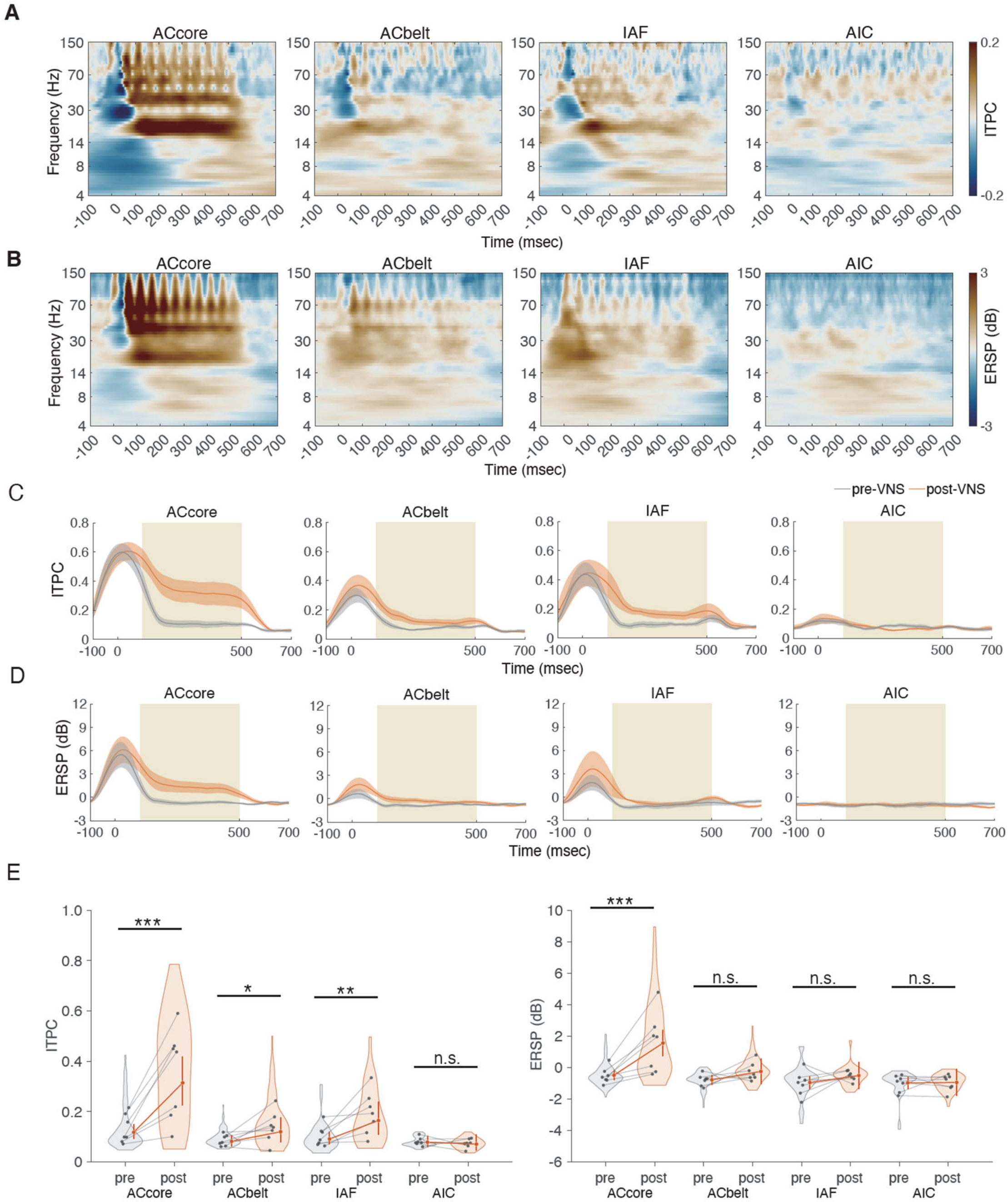

**Fig. 3** Effects of VNS on cortical entrainment to 20-Hz click trains. Layout as in Fig. 2, for 20-Hz ITPC and 20-Hz ERSP. (A) Time–frequency maps of the VNS-induced change in ITPC. (B) Time–frequency maps of the VNS-induced change in ERSP. (C) Time course of 20-Hz ITPC in each region (ACcore, ACbelt, IAF, AIC) before (gray) and after (orange) VNS. Shading around

each trace indicates the SEM across rats. The beige vertical band denotes the 100–500 ms quantification window. (D) Time course of 20-Hz ERSP, displayed as in (C). (E) 20-Hz ITPC (left) and 20-Hz ERSP (right) before and after VNS in each region. $n = 7$ rats. *$p < 0.05$, **$p < 0.01$, ***$p < 0.001$; n.s., not significant.

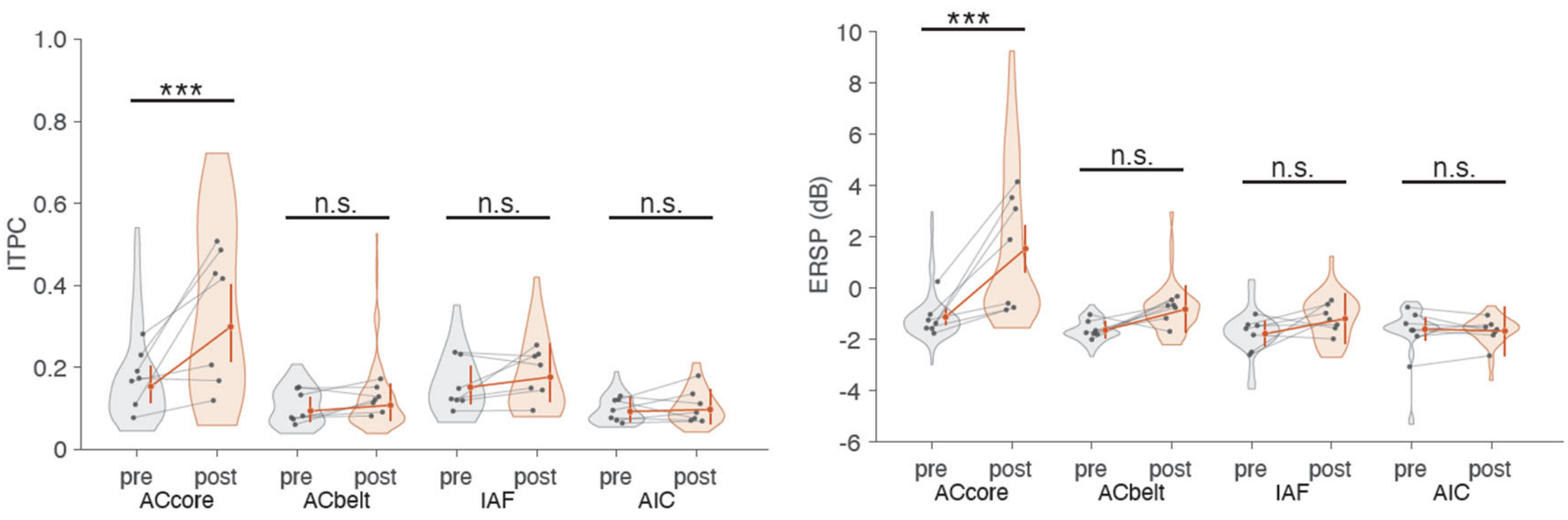

**Fig. 4** Effects of VNS on the 40-Hz second-harmonic response to 20-Hz click trains. Violin plots of 40-Hz ITPC (left) and 40-Hz ERSP (right) before and after VNS in each region, displayed as in Fig. 2E. $n = 7$ rats. ***$p < 0.001$; n.s., not significant.

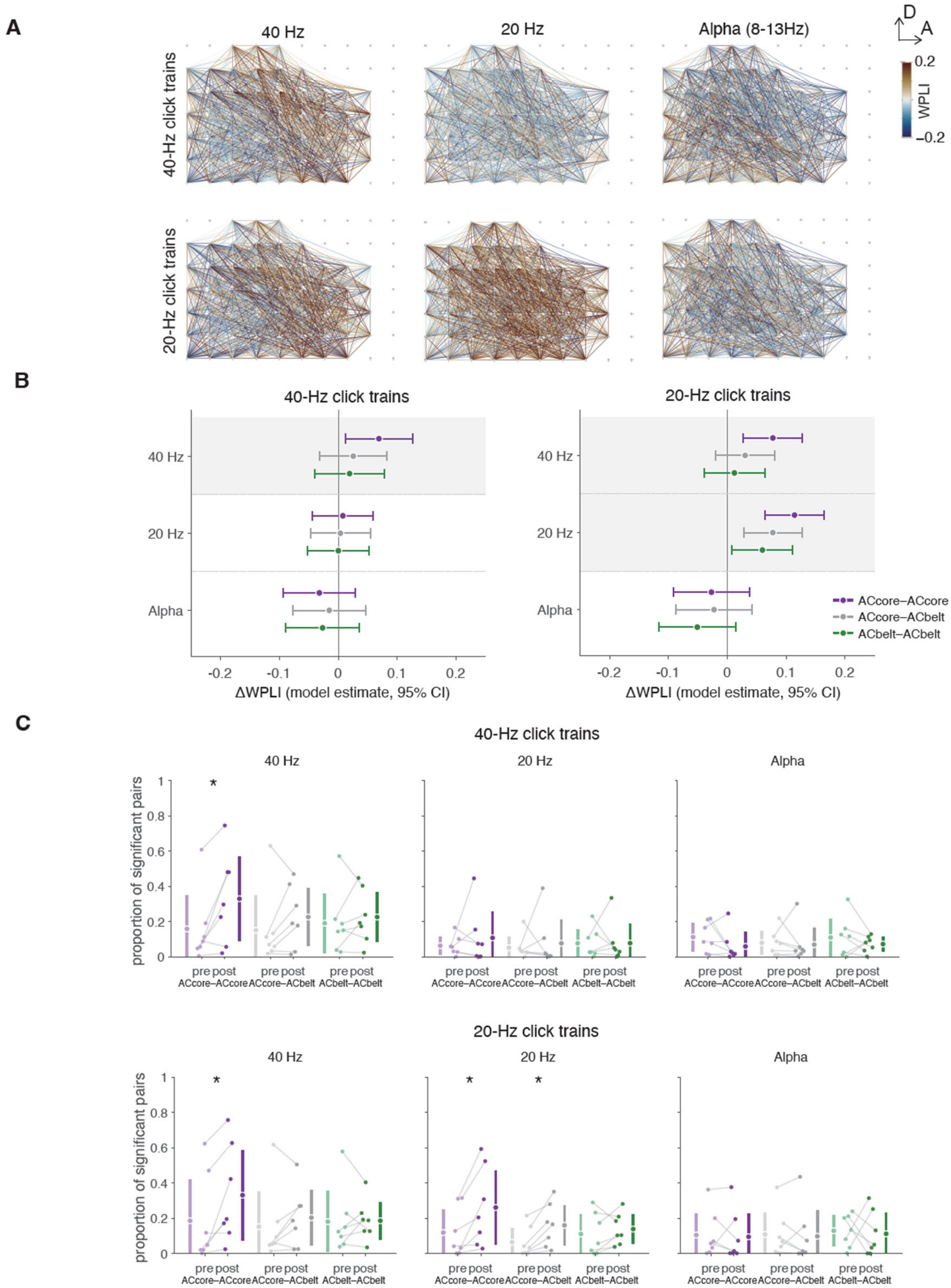

**Fig. 5** Effects of VNS on phase coupling within the auditory cortex. (A) Connectivity maps of the VNS-induced change in WPLI between electrode pairs during 40-Hz (top) and 20-Hz (bottom)

click trains, averaged across rats. Lines connect electrode pairs and are colored by the change in WPLI. All available pairs are shown. (B) Forest plots of the VNS-induced change in WPLI estimated by a linear mixed model, for 40-Hz (left) and 20-Hz (right) click trains. Rows indicate frequency band. Symbols and horizontal lines indicate estimates and 95% confidence intervals for each pair type (purple, ACcore–ACcore; gray, ACcore–ACbelt; green, ACbelt–ACbelt). Gray shading denotes the frequencies present in the click train (40 Hz for 40-Hz trains; 20 and 40 Hz for 20-Hz trains). (C) Proportion of electrode pairs exceeding a surrogate threshold before and after VNS. Symbols indicate means across rats, and error bars indicate 95% confidence intervals. $n = 7$ rats. *$p < 0.05$. WPLI, debiased weighted phase-lag index.

**Supplementary Materials for**

# Vagus nerve stimulation strengthens beta- and gamma-rate auditory steady-state responses in rat cortex

Shinichi Kumagai[a,b], Tomoyo Isoguchi Shiramatsu[b], Karin Oshima[b], Yohei Ishishita[a], Kenji Ibayashi[a], Yoshiyuki Onuki[a], Naoto Kunii[a], Kensuke Kawai[a], Hirokazu Takahashi[b*]

[a]Department of Neurosurgery, Jichi Medical University, Tochigi, Japan; [b]Department of Mechano-Informatics, Graduate School of Information Science and Technology, The University of Tokyo, Tokyo, Japan

Correspondence to: Hirokazu Takahashi (takahashi@i.u-tokyo.ac.jp)

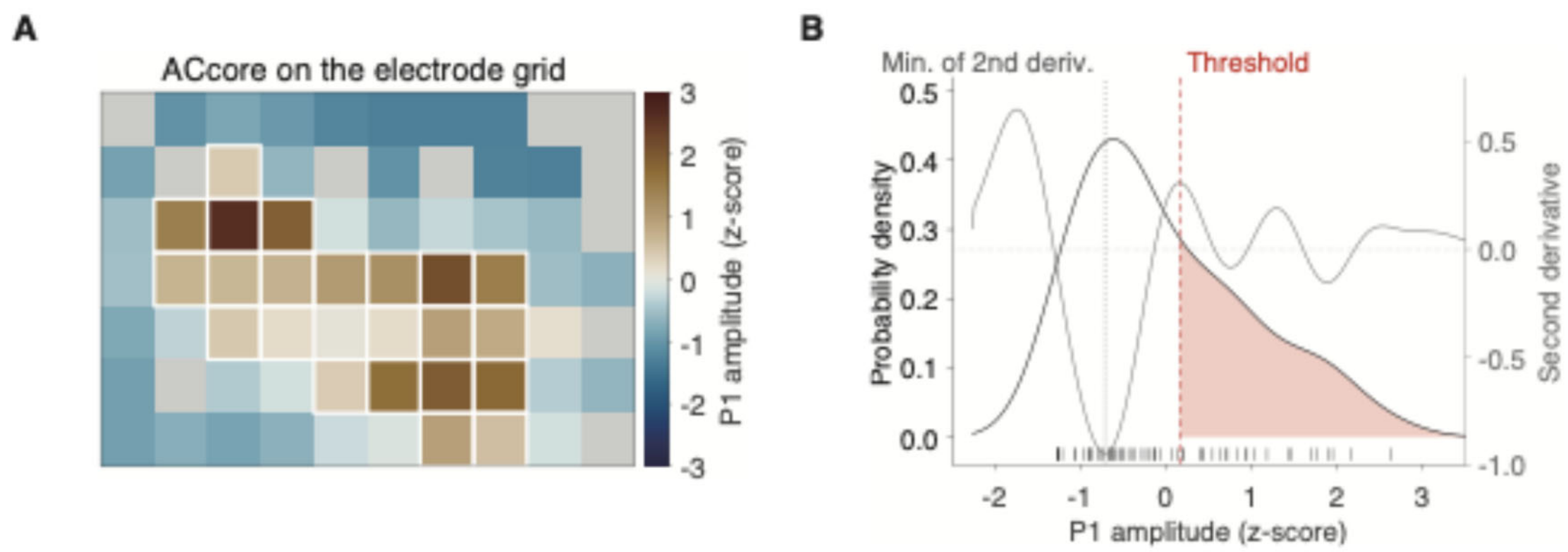

**Supplementary Fig. S1** Delineation of the ACcore in a representative rat. (A) Click-evoked P1 amplitudes, z-scored across the auditory cortical array, displayed on the electrode grid. Gray tiles indicate channels excluded for signal quality or positions without electrodes. White outlines denote sites assigned to the ACcore. (B) Probability density of the z-scored P1 amplitudes estimated with a Gaussian kernel (black) and its second derivative (gray). The minimum of the second derivative was used as a reference point (gray dotted line), and the threshold (red dashed line) was defined as the z score corresponding to the first local maximum of the second derivative toward higher amplitudes. Sites exceeding the threshold were assigned to the ACcore (shaded). Tick marks below the density indicate individual recording sites.

| | Electrodes | | | | | Electrode pairs | | | |
|---|---|---|---|---|---|---|---|---|---|
| **Rat** | **ACcore** | **ACbelt** | **IAF** | **AIC** | **Total** | **ACcore–ACcore** | **ACcore–ACbelt** | **ACbelt–ACbelt** | **Total** |
| **#1** | 9 | 12 | 3 | 6 | 30 | 36 | 108 | 66 | 210 |
| **#2** | 23 | 4 | 6 | 6 | 39 | 253 | 92 | 6 | 351 |
| **#3** | 22 | 24 | 6 | 5 | 57 | 231 | 528 | 276 | 1035 |
| **#4** | 19 | 25 | 3 | 5 | 52 | 171 | 475 | 300 | 946 |
| **#5** | 19 | 17 | 2 | 5 | 43 | 171 | 323 | 136 | 630 |
| **#6** | 22 | 13 | 6 | 6 | 47 | 231 | 286 | 78 | 595 |
| **#7** | 25 | 21 | 7 | 6 | 59 | 300 | 525 | 210 | 1035 |
| **Total** | 139 | 116 | 33 | 39 | 327 | 1393 | 2337 | 1072 | 4802 |

**Supplementary Table S1** Number of electrodes and electrode pairs contributed by each rat. Pair counts are for the auditory array and are identical for the 40- and 20-Hz click conditions.

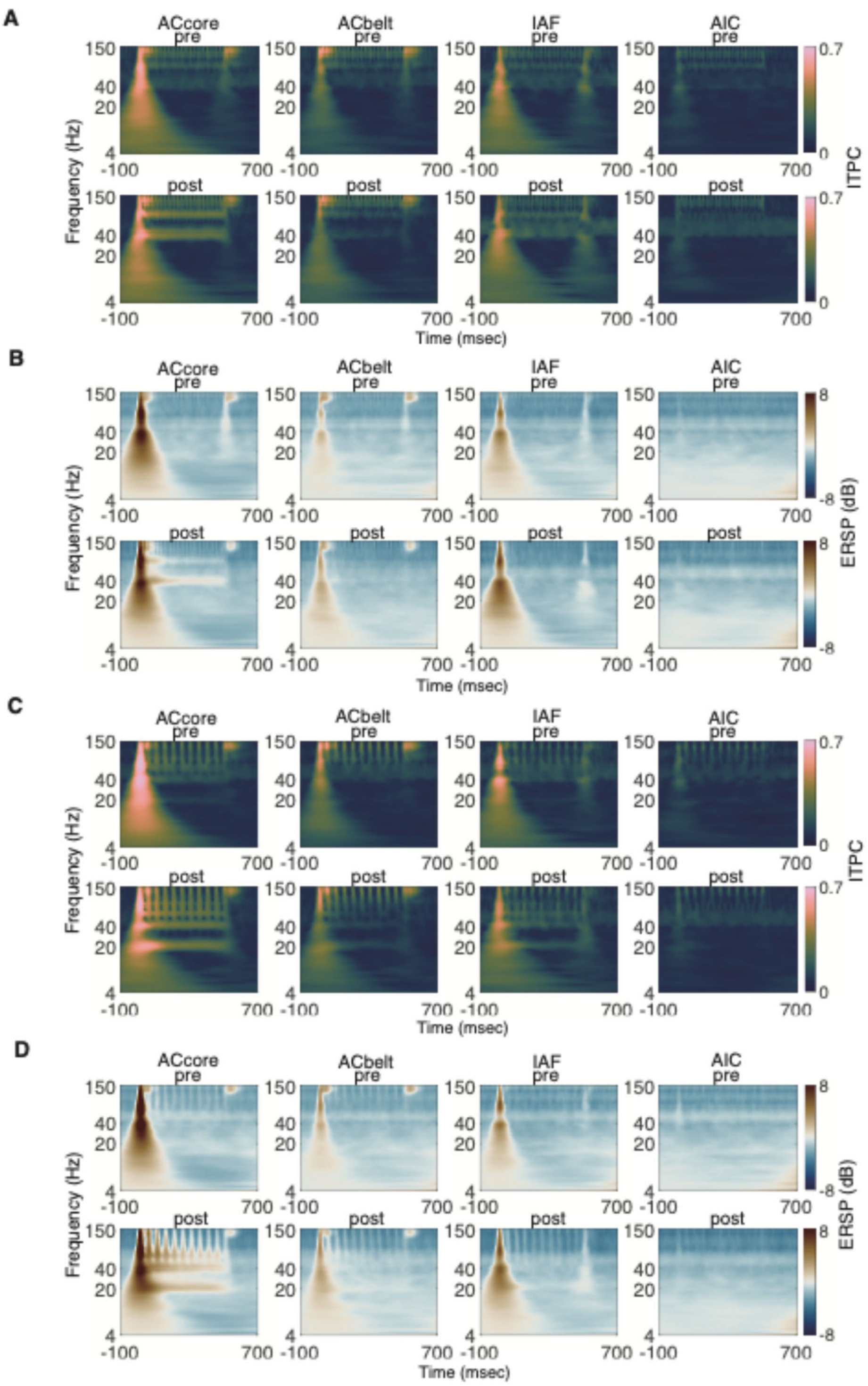

**Supplementary Fig. S2** Time–frequency maps of ITPC and ERSP before and after VNS. (A) ITPC during 40-Hz click trains for each region (ACcore, ACbelt, IAF, AIC) before (pre, upper) and after (post, lower) VNS. Color scale indicates ITPC. (B) ERSP during 40-Hz click trains. Color scale indicates ERSP (dB). (C) ITPC during 20-Hz click trains. (D) ERSP during 20-Hz click trains. Maps show the grand average across 7 rats.

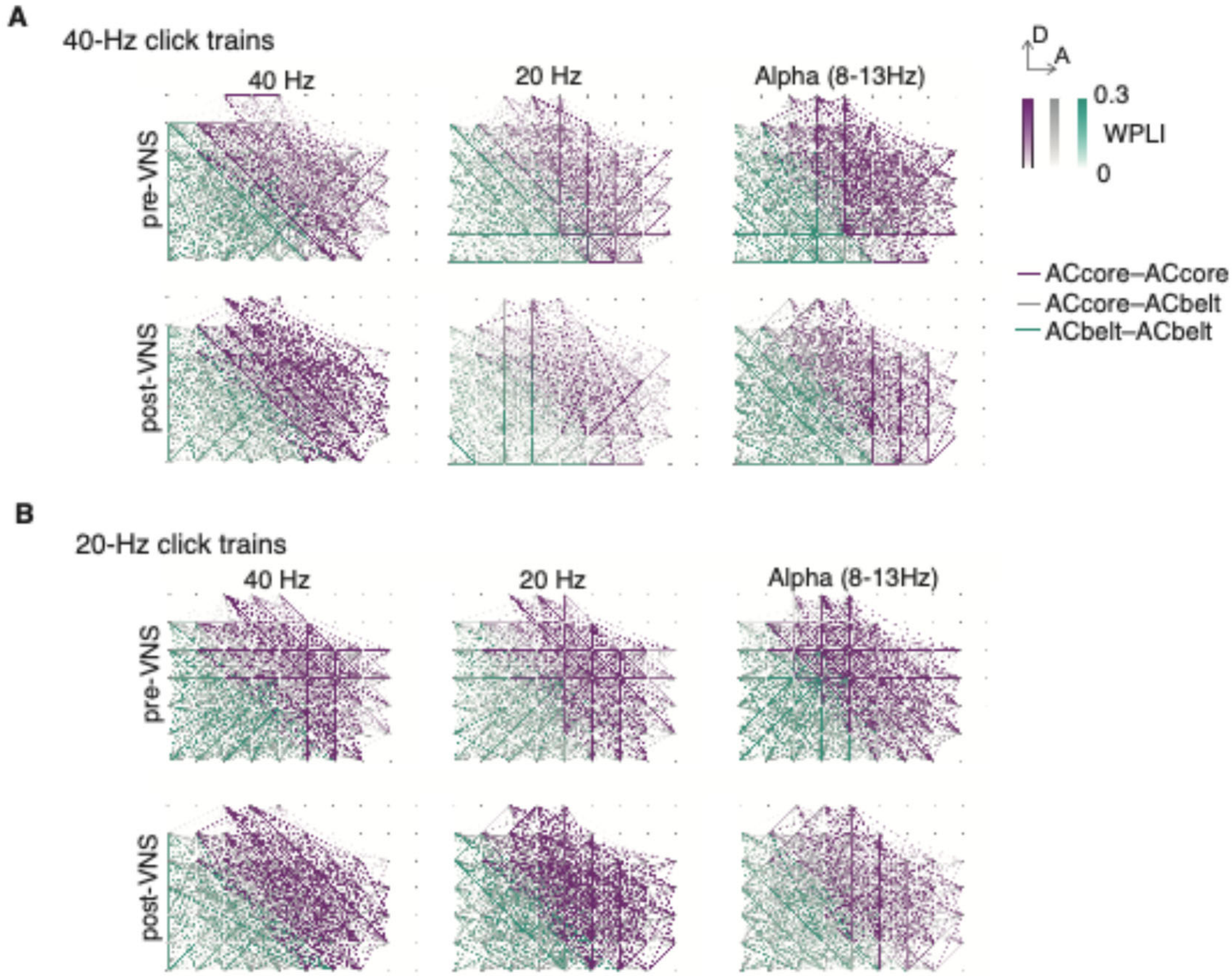

**Supplementary Fig. S3** Connectivity maps of WPLI before and after VNS. (**A**) WPLI between electrode pairs during 40-Hz click trains, before (pre, upper) and after (post, lower) VNS. Each line shows the mean WPLI across the electrode pairs it connects. Color scale indicates WPLI. (**B**) WPLI during 20-Hz click trains, displayed as in (A).

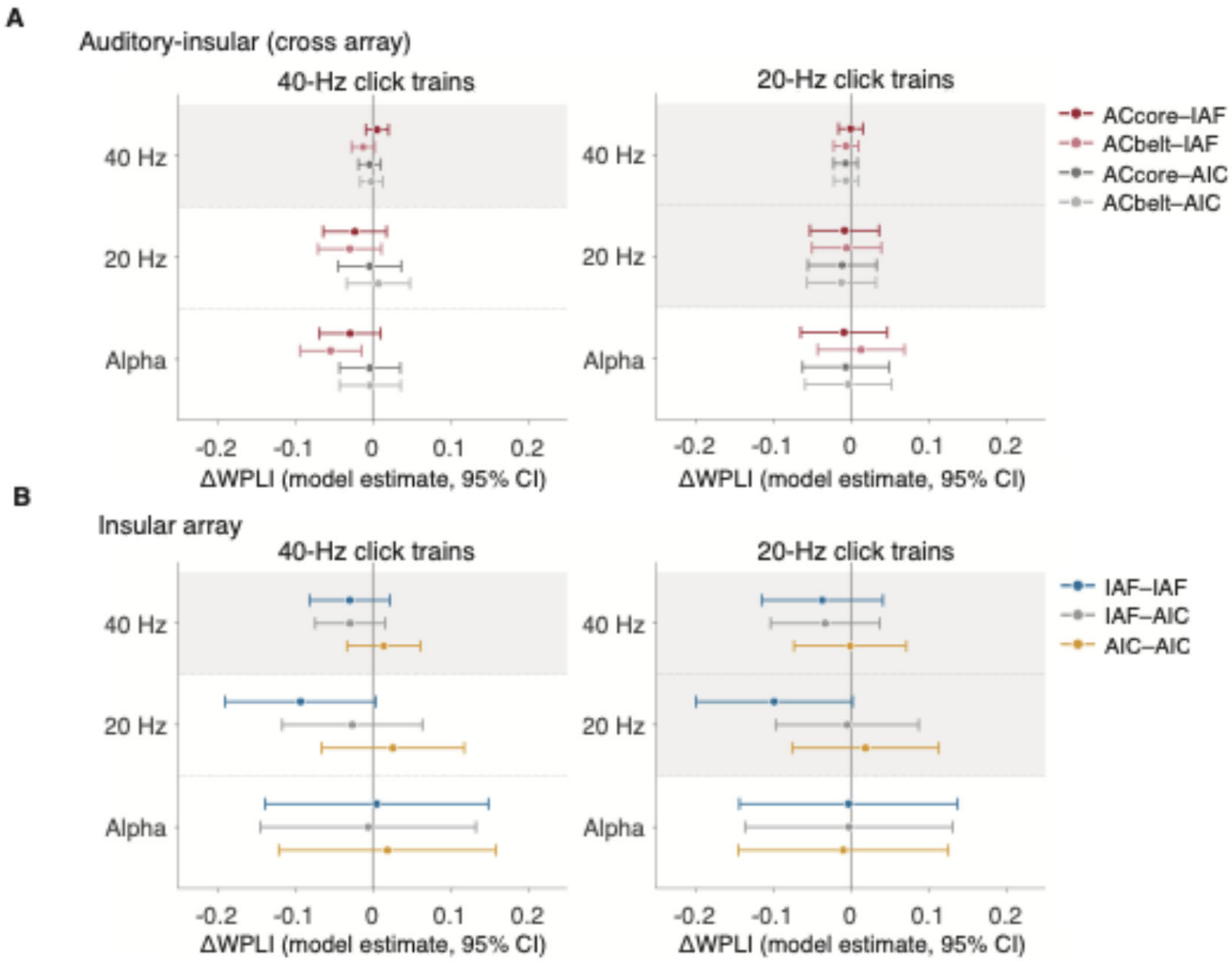

**Supplementary Fig. S4** VNS-induced changes in WPLI for cross-array and insular-array electrode pairs. (A) Forest plots of the change in WPLI estimated by a linear mixed model for cross-array pairs during 40-Hz (left) and 20-Hz (right) click trains. Rows indicate frequency band. Symbols and horizontal lines indicate estimates and 95% confidence intervals for each pair type (dark red, ACcore–IAF; pink, ACbelt–IAF; dark gray, ACcore–AIC; light gray, ACbelt–AIC). Gray shading denotes the driven frequencies (40 Hz for 40-Hz trains; 20 and 40 Hz for 20-Hz trains). The vertical line indicates no change. (B) Forest plots for pairs within the insular array (blue, IAF–IAF; gray, AIC–IAF; yellow, AIC–AIC), displayed as in (A).

## Linear mixed-effects model specification

***P1–N1 amplitude, ITPC, and ERSP***

P1–N1 amplitude, logit-transformed ITPC, and ERSP were analyzed using linear mixed-effects models in SPSS Statistics v31 (IBM). ITPC values never reached 0 or 1, so the logit transformation was applied without a continuity correction. Models were fitted separately by measure; for ITPC and ERSP, separate models were fitted for each click-train condition and response band. Fixed effects were Region, Timing (pre/post), and their interaction. With Post coded 0 at Pre and 1 at Post, the model specification was:

$$y \sim \text{Region*Timing} + (1|\text{rat}) + (0 + \text{Post}|\text{rat:region}) + (1|\text{rat:electrode})$$

The random effects therefore comprised a rat intercept, a Post slope for each region within rat, and an electrode intercept within rat; these were specified as independent variance components (VC). Models were fitted by restricted maximum likelihood (REML), with Satterthwaite denominator degrees of freedom. Benjamini–Hochberg correction was applied separately to the six pairwise regional comparisons and to the four within-region Timing effects, within each measure, click-train condition, and response band.

***WPLI***

The electrode pair was the unit of analysis. WPLI was analyzed untransformed in MATLAB R2022a (fitlme), separately for pairs within the auditory array, across arrays, and within the insular array. Pairs were analyzed in separate families because WPLI and its residual variance depend on electrode configuration. Pair type replaced Region, and centered inter-electrode distance was included for within-array families:

$$y \sim \text{PairType*Timing}\ [+\ \text{Distance_c}] + (1|\text{rat}) + (0 + \text{Post}|\text{rat:PairType})$$
$$+ (1|\text{rat:electrode1}) + (1|\text{rat:electrode2})$$

The distance term was omitted for across-array pairs. The electrode-specific intercepts account for the crossed membership of each pair. These models were fitted in MATLAB rather than SPSS because crossed random effects are not readily specified in SPSS. Models used REML with Satterthwaite degrees of freedom for fixed-effect tests.

***Frequency-specificity models***

The 20- and 40-Hz click-train conditions were combined. A response band was classified as driven when it was present in the click train (40 Hz for 40-Hz trains; 20 and 40 Hz for 20-Hz trains) and as non-driven otherwise (20 Hz during 40-Hz trains and alpha during both train rates). For ITPC and ERSP, fixed effects were Driven status, Timing, their interaction, and Region; the corresponding rat- and electrode-level random intercepts were retained, with the Post slope allowed to vary by rat within each click-rate × response-band combination.

For WPLI, frequency-specificity analyses were restricted to the auditory array and included pair type and centered inter-electrode distance. The primary Driven model was:

```
y ~ Driven*Timing + PairType + Distance_c + (1|rat)
  + (0 + Post|rat:click-rate:band) + (1|rat:electrode1)
  + (1|rat:electrode2)
```

As a complementary analysis, the 20-Hz response band was analyzed separately with:

```
y ~ ClickRate*Timing + PairType + Distance_c + [same random effects]
```

***Model fitting checks***

All models converged without warnings. Estimated random-effect variance components were inspected for boundary estimates; none were zero or near zero.